\documentstyle[12pt]{article}
\begin{document}
\vskip .7cm
\begin{center}
{\bf \Large { New symmetries for Abelian gauge theory\\
in superfield formulation}}

\vskip 2cm

{\bf R.P.Malik}
\footnote{ E-mail address: malik@boson.bose.res.in  }\\
{\it S. N. Bose National Centre for Basic Sciences,} \\
{\it Block-JD, Sector-III, Salt Lake, Calcutta-700 098, India} \\

\vskip 2.5cm

\end{center}

\noindent
{\bf Abstract}:
We show the existence of some new local, covariant and continuous symmetries 
for the BRST invariant Lagrangian density of a free two ($1 + 1$)-dimensional 
(2D) Abelian $U(1)$ gauge theory in the framework of superfield formalism. The 
Noether conserved charges corresponding to the above local 
continuous symmetries find their geometrical origin
as the translation generators along the odd (Grassmannian)- and even (bosonic) 
directions of the four ($2 + 2)$-dimensional compact supermanifold.  
Some new discrete symmetries are shown to exist in the superfield formulation. 
The logical origin for the existence of BRST- and co-BRST
symmetries is shown to be encoded in the Hodge decomposed
versions (of the 2D fermionic vector fields) that are consistent with
the discrete symmetries of the theory.
\baselineskip=16pt


\newpage

\noindent
{\bf 1 Introduction}\\

\noindent
The superfield approach [1--5] to Becchi-Rouet-Stora-Tyutin (BRST) 
formalism is a well-established
technique which provides the geometrical origin for the existence of
(anti-)BRST charges as the generators of translation along the Grassmannian
directions of the compact supermanifold that is parametrized by the spacetime
coordinates and two extra anti-commuting (Grassmannian) variables. In fact,
in this scheme, the ($ p + 1$)-form super curvature tensor 
for a $p$-form ($ p = 1, 2, 3,...$) gauge theory is restricted to
be flat along the Grassmannian directions of the supermanifold. This 
restriction, popularly known as the horizontality condition
\footnote{This condition is referred to as the ``soul flatness'' condition
in Ref. [6] implying the flatness of the Grassmannian components of the
($ p + 1$-form) super curvature tensor for a $p$-form gauge theory.},
leads to the derivation of the (anti-)BRST symmetry transformations for
the Lagrangian density of a $p$-form gauge theory. In this derivation,
the mathematical power of the super exterior derivative $\tilde d$ 
{\it alone} (which is only one of 
the {\it three} de Rham cohomology operators 
\footnote{On an ordinary Minkowskian manifold parametrized by the spcetime
co-ordinate $x^\mu$, the exterior derivative $d$ ($d = dx^\mu \partial_{\mu},
d^2 = 0$), the co-exterior derivative $\delta$ ($\delta = \pm * \;d\;*;
\delta^2 = 0, * = $ Hodge duality operation) and the Laplacian operator
$\Delta (\Delta = (d + \delta)^2 = d \delta + \delta d)$ constitute 
what is popularly known as the set $(d, \delta, \Delta)$  of the de Rham 
cohomology operators. These geometrical operators obey: 
$ \delta^2 = 0, d^2 = 0, \{ d , \delta \} = \Delta, [\Delta , d] =
[\Delta, \delta ] = 0$ implying that $\Delta$ is the Casimir
operator for this algebra [7-10].} of differential geometry)  is
exploited when it operates on the super $p$-form potential of a $p$-form
gauge theory to make it a $(p + 1)$-form curvature tensor through Maurer-Cartan
equation. Thus, it is
an interesting endeavour to explore the possibility of the existence of some
new local symmetries by exploiting the other two super de Rham cohomology 
operators ($\tilde \delta$: co-exterior derivative;
$\tilde \Delta$: Laplacian operator) of differential geometry
and find out their geometrical interpretation in the language
of some kind of translation generators
on an appropriately chosen compact supermanifold.

The purpose of the present paper is to show the existence of some new local,
covariant and continuous symmetries for the free 2D Abelian gauge theory that
emerge due to the operation of super co-exterior derivative $\tilde \delta$
($\tilde \delta = - \tilde * \tilde d \tilde *, \tilde \delta^2 = 0, \tilde
* = $ Hodge duality operation)
and super Laplacian operator $\tilde\Delta$ ($\tilde \Delta = \tilde d \tilde 
\delta + \tilde \delta \tilde d$) on the super one-form connection $\tilde A$
together with the analogue of the horizontality conditions w.r.t. these
super de Rham cohomology operators. In fact, we demonstrate that
(anti-)co-BRST symmetry- and a bosonic symmetry transformations
emerge when we exploit these super cohomological operators on a 
four ($2 + 2$)-dimensional compact supermanifold and they turn
out to be exactly same as the new local symmetries obtained recently
in a set of papers in the Lagrangian formalism {\it alone} [11-15]. It has been 
established in these works that the 2D free- as well
as interacting (non-)Abelian (one-form) gauge theories provide the field 
theoretical models for the Hodge theory where all the de Rham cohomology 
operators find their interpretation as the local Noether charges that generate 
these new local, continuous and covariant symmetries. Such symmetries and 
corresponding generators (conserved Noether charges)
have also been shown to exist for the 
four ($3 + 1$)-dimensional free two-form Abelian gauge theory [16]. In these
attempts, the local Noether charges have also been shown to refine the
BRST cohomology [12] and define the analogue of the Hodge decomposition 
theorem(HDT)
\footnote{ This theorem states that, on a compact manifold without a boundary, 
any arbitrary $n$-form $f_{n}, (n = 0, 1, 2..)$ can be uniquely written as 
the sum of a harmonic form $h_{n} (\Delta h_{n} = d h_{n} = \delta h_{n} = 0)$,
an exact form $d e_{n-1}$ and a co-exact form $\delta c_{n+1}$ 
as: $f_{n} = h_{n} + d e_{n-1} + \delta c_{n+1}$ [7-10].}
in the quantum Hilbert space of states [11-17]. 
Exploiting these ideas, it has been shown that 2D free (non-)Abelian gauge 
theories belong to a new class of topological field theories (TFTs) [17]. 
However, in all the above attempts, the geometrical origin for the 
existence of these charges has {\it not yet} been discussed. In the present 
work, we show that the (anti-)BRST- and (anti-)co-BRST
symmetry generators (conserved and nilpotent Noether charges
$(\bar Q_{b})Q_{b}$ and $(\bar Q_{d})Q_{d}$ respectively) are the
translation generators along the Grassmannian (odd) directions of the
$(2 + 2)$-dimensional compact supermanifold and they owe their origin 
to the super cohomological operators $\tilde d$ and $\tilde \delta$. 
A bosonic symmetry, generated by the Casimir operator, turns out to 
be the translation generator along the bosonic (even) direction of the 
supermanifold and its origin is encoded in the super operator
$\tilde \Delta$. This even (bosonic) direction on the supermanifold is 
equivalent to a couple of intertwined Grassmannian directions. The local 
conserved charges in the theory provide an analogue of the 
set $(d, \delta, \Delta)$.

The outline of our present paper is as follows. In Sec. 2, we set up the
notations and recapitulate some of the salient features of our earlier works
[11-17]. Section 3 is devoted to the derivation of (anti-)BRST symmetry
transformations through horizontality condition [3,4]. In Sec. 4, we exploit 
the super co-exterior derivative and derive the (anti-)co-BRST symmetry 
transformations exploiting the analogue of the horizontality condition
w.r.t. $\tilde \delta$. We discuss some interesting discrete symmetries 
and the Hodge decomposed versions of 2D vectors in Sec. 5.  A local bosonic 
symmetry is obtained in Sec. 6 using the super Laplacian operator 
$\tilde \Delta$. Finally, we make some concluding remarks in Sec. 7.\\

\noindent
{\bf 2 BRST- and dual BRST symmetries: A brief sketch}\\

\noindent
Let us start off with the BRST invariant Lagrangian density ${\cal L}_{b}$
for the free two ($1 + 1)$-dimensional
\footnote{We follow here the conventions and notations such that the 2D flat
Minkowski metric is: $\eta_{\mu\nu} =$ diag $(+1, -1)$ and $\Box = 
\eta^{\mu\nu} \partial_{\mu} \partial_{\nu} = (\partial_{0})^2 - 
(\partial_{1})^2, \varepsilon_{\mu\nu} = - \varepsilon^{\mu\nu}, F_{01} = E
= - \varepsilon^{\mu\nu} \partial_{\mu} A_{\nu} = 
\partial_{0} A_{1} - \partial_{1} A_{0}= F^{10}, \varepsilon_{01} =
\varepsilon^{10} = + 1.$ Here the Greek indices: $\mu, \nu, \lambda...= 0, 1$ 
correspond to spacetime directions on the 2D manifold.} 
Abelian gauge theory in the Feynman gauge [6,18-20]
$$
\begin{array}{lcl}
{\cal L}_{b} = - \frac{1}{4}\; F^{\mu\nu} F_{\mu\nu} 
- \frac{1}{2} (\partial \cdot A)^2 - i \partial_{\mu} \bar C \partial^\mu C 
\equiv \frac{1}{2}\; E^2 - \frac{1}{2} (\partial
\cdot A)^2 - i \partial_{\mu} \bar C \partial^\mu C, 
\end{array} \eqno(2.1)
$$
where $F_{\mu\nu} = \partial_{\mu} A_{\nu} - \partial_{\nu} A_{\mu}$ is the
field strength tensor (curvature two-form) derived from one-form $A = dx^\mu
A_{\mu}$ (with $A_{\mu} =$ vector potential) by the application of the exterior
derivative $d $ (i.e. $ F = d A 
=\frac{1}{2} d x^\mu \wedge d x^\nu F_{\mu\nu}$). The gauge-fixing term 
(zero-form) is derived from one-form $A$ by the application of the co-exterior
derivative $\delta$ (i.e. $(\partial \cdot A) = \delta A; \delta = - * d *;
* =$ Hodge duality operation). Thus, in some sense,
$F = d A$ and $(\partial \cdot A) 
= \delta A$ are ``Hodge dual'' to each-other. The (anti-)ghost fields ($\bar
C) C$ are anti-commuting ($ C^2 = \bar C^2 = 0, C \bar C + \bar C C = 0$)
in nature. Under the following on-shell ($ \Box C = \Box \bar C = 0$) nilpotent
$(s_{b}^2 = 0, \bar s_{b}^2 = 0, s_{b} \bar s_{b} + \bar s_{b} s_{b} = 0)$ 
(anti-)BRST transformations ($\bar s_{b}) s_{b}$
\footnote{Here the notations, followed in Ref. [20], are adopted. In fact,
in its totality, a BRST transformation $\delta_{B}$ is the product of an 
anti-commuting spacetime independent parameter $\eta$ and $s_{b}$ 
(i.e. $\delta_{B} = \eta \; s_{b}$).} on the basic fields:
$$
\begin{array}{lcl}
s_{b} A_{\mu} &=& \partial_{\mu} C, \qquad s_{b} C = 0, \qquad 
s_{b} \bar C = -\; i\; (\partial \cdot A), \nonumber\\
\bar s_{b} A_{\mu} 
&=& \partial_{\mu} \bar C, \qquad \bar s_{b} \bar C = 0, \qquad 
\bar s_{b}  C = +\; i\; (\partial \cdot A), 
\end{array} \eqno(2.2)
$$
the Lagrangian density (2.1) remains invariant. The same Lagrangian density
is also invariant under the following on-shell ($\Box C = \Box \bar C = 0$)
nilpotent ($s_{d}^2 = \bar s_{d}^2 = 0,\; s_{d} \bar s_{d} + \bar s_{d} s_{d}
= 0$) (anti-)dual BRST transformations $(\bar s_{d})s_{d}$ on the basic
fields [11,12,17]
$$
\begin{array}{lcl}
s_{d} A_{\mu} &=& - \varepsilon_{\mu\nu} \partial^\nu \bar C, \qquad
s_{d} \bar C = 0, \qquad s_{d} C = - i E, \nonumber\\
\bar s_{d} A_{\mu} &=& - \varepsilon_{\mu\nu} \partial^\nu C, \qquad
\bar s_{d}  C = 0, \qquad \bar s_{d} \bar C = + i E. 
\end{array} \eqno(2.3)
$$
We christen the above new continuous, covariant and nilpotent symmetry as
the (anti-) dual BRST symmetry because it is the gauge-fixing term $(\partial
\cdot A) = \delta A$ (Hodge dual to the curvature two-form
$F = d A$) that remains invariant. In contrast, it is the curvature two-form
$F = d A$ that remains invariant under the (anti-)BRST symmetry 
transformations (2.2). The anti-commutator of the above two symmetries
leads to yet another new bosonic type symmetry transformation $s_{w}$
$(s_{w}= \{ s_{b}, s_{d} \} = \{ \bar s_{b}, \bar s_{d} \}; s_{w}^2 \neq 0)$
[12]
$$ 
\begin{array}{lcl}
s_{w} A_{\mu} =  \partial_{\mu} E
- \varepsilon_{\mu\nu} \partial^\nu (\partial \cdot A) \equiv
- \varepsilon_{\mu\nu} \Box A^{\nu}, \qquad s_{w} C = 0, 
\qquad s_{w} \bar C = 0,
\end{array} \eqno(2.4)
$$
under which the ghost fields remain invariant. The Noether 
conserved charges ($Q_{r}$) corresponding to the above continuous symmetries
are the generators for the above transformations [11-17]. This statement
can be concisely expressed as
$$
\begin{array}{lcl}
s_{r} \Psi = - i \; [ \Psi, Q_{r} ]_{\pm}, \qquad \;\;\;
Q_{r} = Q_{b}, \bar Q_{b}, Q_{d}, \bar Q_{d}, Q_{w}, Q_{g}, 
\end{array} \eqno(2.5)
$$
where brackets $[\;, \;]_{\pm}$ stand for the (anti-)commutators for 
any arbitrary generic field $\Psi$ being (fermionic)bosonic
in nature. Here the ghost charge $Q_{g}$ ($Q_{g} = - i \int dx [C \dot {\bar C}
+ \bar C \dot C]$) generates
the continuous scale transformations: $ C \rightarrow e^{-\Sigma} C,
\bar C \rightarrow e^{\Sigma} \bar C, A_{\mu} \rightarrow A_{\mu}$ (where
$\Sigma$ is a global parameter) for the invariance of the Lagrangian
density (2.1).

Now we wish to discuss some of the discrete symmetries 
present in the theory. It is 
interesting to note that $s_{b} \leftrightarrow \bar s_{b}$ under the
discrete symmetry transformations : $ C \leftrightarrow \bar C,\; 
(\partial \cdot A) \leftrightarrow - (\partial \cdot A)$. On the other hand, 
$s_{d} \leftrightarrow
\bar s_{d}$ when we take: $ C \leftrightarrow \bar C,\; E \leftrightarrow - E$.
Under yet another discrete symmetry transformations [17,15]
$$
\begin{array}{lcl}
C\rightarrow \pm i \bar C,\; \bar C \rightarrow \pm i C, \;
E \rightarrow \pm i (\partial \cdot A),\; (\partial \cdot A) \rightarrow 
\pm i E, \; A_{\mu} \rightarrow A_{\mu}, \partial_{\mu} \rightarrow \pm i
\varepsilon_{\mu\nu} \partial^\nu,
\end{array} \eqno(2.6)
$$
the Lagrangian density (2.1) remains form-invariant and the symmetry
transformations (2.2) and (2.3) are related to one-another. Furthermore,
this discrete symmetry turns out to be the analogue of the Hodge $*$ duality
operations of the differential geometry as one of the key relationships:
$ s_{d}\; \Psi = \pm \; *\; s_{b}\; * \Psi, (\bar s_{d}\; \Psi 
= \pm \; *\; \bar s_{b}\; * \Psi)$ 
exists for any arbitrary generic 
field $\Psi$ of the theory [15,17]. The ($\pm$) sign in this
relationship is dictated by the existence of the corresponding sign in the
operation: $ *\;(\; *\; \Psi) = \pm\; \Psi$ where $*$ is nothing but the
discrete transformations (2.6). The Lagrangian density (2.1)
and the corresponding symmetric energy-momentum tensor $T_{\mu\nu}$
can be expressed, modulo some total derivatives, as [11,12,17] 
$$
\begin{array}{lcl}
{\cal L}_{b} &=& \{ Q_{d}, S_{1}\} + \{ Q_{b}, S_{2}\} \equiv
s_{d}\; (i S_{1}) + s_{b}\; (i S_{2}), \nonumber\\
T_{\mu\nu} &=& \{ Q_{d}, V^{(1)}_{\mu\nu}\} + \{ Q_{b}, V^{(2)}_{\mu\nu}\}
\equiv s_{d}\; (i V^{(1)}_{\mu\nu}) + s_{b}\; (i V^{(2)}_{\mu\nu}), 
\end{array} \eqno(2.7)
$$
where $ S_{1} = \frac{1}{2}\; E C, \; S_{2} = - \frac{1}{2}\; 
(\partial \cdot A) \bar C$ and the local field dependent expressions for
$V's$ are
$$
\begin{array}{lcl}
V^{(1)}_{\mu\nu} 
&=& \frac{1}{2} \bigl [\;(\partial_{\mu} C)\; \varepsilon_{\nu\lambda}
A^\lambda + (\partial_{\nu} C)\; \varepsilon_{\mu\lambda} A^\lambda -
\eta_{\mu\nu}\; E C\; \bigr ],\nonumber\\
V^{(2)}_{\mu\nu} &=& \frac{1}{2} \bigl [\;(\partial_{\mu} \bar C) \;
A_\nu + (\partial_{\nu} \bar C)\;  A_\mu  +
\eta_{\mu\nu}\; (\partial \cdot A)\; \bar C \;\;\bigr ].
\end{array} \eqno(2.8)
$$
The expressions in (2.7) establish the topological nature of 2D free Abelian
gauge theory as topological invariants and their recursion relations have
been obtained in Ref. [17]. The algebra amongst the conserved charges
of the theory are reminiscent of the algebra obeyed by the de Rham cohomology
operators of differential geometry. Thus, the present theory
is a field theoretic  model for the Hodge theory and it represents a new
class of topological field theory which captures some of the key features
of Witten- and Schwarz type TFTs.\\

\noindent
{\bf 3 Super exterior derivative and (anti-)BRST symmetry transformations}\\

\noindent
We begin with the definition of a super exterior derivative ($\tilde d$)
and a super one-form connection ($\tilde A$) on a $( 2 + 2)$-dimensional
compact supermanifold as [21]
$$
\begin{array}{lcl}
\tilde d &=& \;d Z^M \;\partial_{M} = d x^\mu\; \partial_\mu\;
+ \;d \theta \;\partial_{\theta}\; + \;d \bar \theta \;\partial_{\bar \theta},
\nonumber\\
\tilde A &=& d Z^M\; \tilde A_{M} = d x^\mu \;B_{\mu} (x , \theta, \bar \theta)
+ d \theta\; \bar \Phi (x, \theta, \bar \theta) + d \bar \theta\;
\Phi ( x, \theta, \bar \theta),
\end{array}\eqno(3.1)
$$
where supermanifold is parametrized by the superspace coordinates 
$Z^M = (x^\mu, \theta, \bar \theta)$ with two c-number (commuting) spacetime
co-ordinates $x^\mu$ (with $\mu = 0, 1)$ and two Grassmann (anti-commuting)
variables $\theta$ and $\bar \theta$ (with $ \theta^2 = \bar \theta^2 = 0,
\theta \bar \theta + \bar \theta \theta = 0)$ and partial derivatives, 
with respect to these superspace coordinates, are
$$
\begin{array}{lcl}
\partial_{M} = {\displaystyle \frac{\partial}{\partial Z^M}},\quad
\partial_{\mu} = {\displaystyle \frac{\partial}{\partial x^\mu}},\quad
\partial_{\theta} = {\displaystyle \frac{\partial}{\partial \theta}},\quad
\partial_{\bar \theta} = {\displaystyle \frac{\partial}{\partial \bar \theta}}.
\end{array}\eqno (3.2)
$$
The bosonic (commuting) superfield $B_{\mu} (x, \theta, \bar \theta)$ and
the fermionic (anti-commuting) superfields: $\Phi (x, \theta, \bar \theta),
\bar \Phi (x, \theta, \bar \theta)$, constitute the component multiplet of
a supervector superfield $V_{s}$, defined on the four-dimensional compact
supermanifold, as [3,4]
$$
\begin{array}{lcl}
V_{s} = \Bigl ( B_{\mu} (x, \theta, \bar \theta), 
\;\Phi (x, \theta, \bar \theta),
\;\bar \Phi (x, \theta, \bar \theta) \Bigr ).
\end{array} \eqno(3.3)
$$
The above superfields
can be expanded in terms of the superspace coordinates $(x^\mu, \theta, 
\bar \theta)$, the field variables of the Lagrangian density (2.1) and
some extra (secondary) fields, as
$$
\begin{array}{lcl}
B_{\mu}\; (x, \theta, \bar \theta) &=& A_{\mu} (x) 
+ \;\theta\; \bar R_{\mu} (x) + \;\bar \theta\; R_{\mu} (x) 
+ \;i \;\theta \;\bar \theta S_{\mu} (x), \nonumber\\
\Phi\; (x, \theta, \bar \theta) &=& C (x) + \;i\;
 \theta\; (\partial \cdot A) (x)
-\; i \;\bar \theta\; E (x) +\; i\; \theta\; \bar \theta \;s (x), \nonumber\\
\bar \Phi\; (x, \theta, \bar \theta) &=& \bar C (x) 
+\; i \;\theta\; E (x) -\; i\; \bar \theta \;(\partial \cdot A)(x) 
+ \;i \;\theta \;\bar \theta \;\bar s (x). 
\end{array} \eqno(3.4)
$$
Here the signs in the expansion are chosen for the later convenience. It is
straightforward to see that the local fields $ R_{\mu}(x), \bar R_{\mu} (x),
C(x), \bar C (x), s (x), \bar s(x)$ are fermionic (anti-commuting) in nature
and the bosonic (commuting) local fields are: $A_{\mu} (x), S_{\mu} (x),
\pm E(x), \pm (\partial \cdot A) (x)$ in the above expansion so that bosonic-
 and fermionic degrees of freedom can match. It is interesting to note that
the above expansion is such that: $(\Phi (x, \theta, \bar \theta))^2 = 0,
(\bar \Phi (x, \theta, \bar \theta))^2 = 0,\; \Phi (x, \theta, \bar \theta)\;
\bar \Phi (x, \theta, \bar \theta) + \bar \Phi (x, \theta, \bar \theta)\;
\Phi (x, \theta, \bar \theta) = 0$ and $ [ B_{\mu} (x, \theta, \bar \theta),
B_{\nu} (x, \theta, \bar \theta) ] = 0$. As a consequence, it is 
straightforward
to verify
that $ \tilde A \wedge \tilde A = \frac{1}{2} [ \tilde A, \tilde A] = 0$.

The super curvature tensor (two-form $\tilde F$) for the  
gauge theory can be constructed by exploiting (3.1) (i.e. $\tilde F = 
\tilde d \; \tilde A + \tilde A \wedge \tilde A$). For the $U(1)$ gauge
theory
$$
\begin{array}{lcl}
\tilde F &=& \tilde d \tilde A = (d x^\mu \wedge d x^\nu)\;
(\partial_{\mu} B_\nu) - (d \theta \wedge d \theta)\; (\partial_{\theta}
\bar \Phi) + (d x^\mu \wedge d \bar \theta)
(\partial_{\mu} \Phi - \partial_{\bar \theta} B_{\mu}) \nonumber\\
&-& (d \theta \wedge d \bar \theta) (\partial_{\theta} \Phi 
+ \partial_{\bar \theta} \bar \Phi) 
+ (d x^\mu \wedge d \theta) (\partial_{\mu} \bar \Phi - \partial_{\theta}
B_{\mu}) - (d \bar \theta \wedge d \bar \theta)
(\partial_{\bar \theta} \Phi), 
\end{array}\eqno(3.5)
$$
where use has been made of the fact that the nilpotency of the super exterior
derivative ($\tilde d ^2 = 0$) implies the following relations for the wedge
products on the supermanifold $ (d x^\mu \wedge d x^\nu) 
= - (d x^\nu \wedge d x^\mu), (d x^\mu \wedge d \theta) 
= - (d \theta \wedge d x^\mu), 
(d \theta \wedge d \bar \theta) = + (d \bar \theta \wedge d \theta)
$ etc.  Now the soul-flatness (or horizontality) condition imposes the
following restriction
$$
\begin{array}{lcl} 
\tilde F = \tilde d \tilde A = \frac{1}{2}\; (d Z^M \wedge d Z^N)\;
\tilde F_{MN} \equiv F = d A = \frac{1}{2}\; (d x^\mu \wedge d x^\nu)\;
F_{\mu\nu}.
\end{array} \eqno(3.6)
$$
In the language of the component superfields of (3.3), 
the above condition implies 
$$
\begin{array}{lcl}
\partial_{\theta}\;\bar \Phi = 0, \quad \partial_{\bar \theta} \Phi = 0,
\quad \partial_{\theta} \Phi + \partial_{\bar \theta} \bar \Phi = 0,\quad
\partial_{\mu} \bar \Phi = \partial_{\theta} B_{\mu}, \quad
\partial_{\mu} \Phi = \partial_{\bar \theta} B_{\mu},
\end{array} \eqno(3.7)
$$
and the following conditions on the local component fields of the superfield
$B_{\mu} (x, \theta, \bar \theta)$
$$
\begin{array}{lcl}
\partial_{\mu} \bar R_{\nu} - \partial_{\nu} \bar R_{\mu} = 0, \quad
\partial_{\mu}  R_{\nu} - \partial_{\nu}  R_{\mu} = 0, \quad
\partial_{\mu} S_{\nu} - \partial_{\nu} S_{\mu} = 0.
\end{array} \eqno(3.8)
$$
The conditions (3.7) lead to the following solutions
$$
\begin{array}{lcl}
R_{\mu} \;(x) &=& \partial_{\mu}\; C(x), \qquad 
\bar R_{\mu}\; (x) = \partial_{\mu}\;
\bar C (x), \qquad \;s\; (x) = 0,
\nonumber\\
S_{\mu}\; (x) &=& - \partial_{\mu}\; (\partial \cdot A) (x),
\;\; \qquad \bar s\;(x) = 0, \;\qquad E\;(x) = 0.
\end{array} \eqno(3.9)
$$
It will be noticed that the signs in the expansion (3.4) are chosen such that
the condition: $ \partial_{\theta} \Phi + \partial_{\bar \theta} \bar \Phi
 = 0$ is satisfied trivially. Furthermore, the solutions in (3.9) 
automatically satisfy the conditions in (3.8). Now the expansion in (3.4)
can be expressed as
$$
\begin{array}{lcl}
B_{\mu}\; (x, \theta, \bar \theta) &=& A_{\mu} (x) 
+ \;\theta\; (\bar s_{b} A_{\mu} (x)) + \;\bar \theta\; (s_{b} A_{\mu} (x)) 
+ \;\theta \;\bar \theta (s_{b} \bar s_{b} A_{\mu} (x)), \nonumber\\
\Phi\; (x, \theta, \bar \theta) &=& C (x) + \; \theta\; (\bar s_{b} C (x))
, \nonumber\\
\bar \Phi\; (x, \theta, \bar \theta) &=& \bar C (x) 
+\bar \theta\; (s_{b} \bar C (x)).
\end{array} \eqno(3.10)
$$
We conclude that the horizontality condition on the super two-form
curvature tensor for the $U(1)$ Abelian gauge theory leads to the derivation
of BRST- and anti-BRST symmetries for the Lagrangian density (2.1). The
corresponding conserved and nilpotent charges find their geometrical origin
as the translation generators along the Grassmannian directions of the
supermanifold. In other words, it is the power of 
$\tilde d$ that provides the geometrical interpretation for $Q_{b}$ and
$\bar Q_{b}$ as translation generators (cf. (2.5)). Thus, the mapping is:
$ \tilde d \Leftrightarrow (Q_{b}, \bar Q_{b})$ but the ordinary exterior
derivative $d$ is identified with $Q_{b}$ {\it alone} because the latter
increases the ghost number of a state by one [11,12,17] as $d$ increases the 
degree of a form by one on which it operates [7-10].\\

\noindent
{\bf 4 Super co-exterior derivative and (anti-)co-BRST 
symmetry transformations}\\

\noindent
We operate the super co-exterior derivative
$ \tilde \delta = - \tilde *\; \tilde d\; \tilde *$ on the super one-form
connection $\tilde A$ of (3.1), with the Hodge duality operation $\tilde *$
defined on the differentials and their wedge products (for the case of
$(2 + 2)$-dimensional compact supermanifold), as 
$$
\begin{array}{lcl}
\tilde *\;\; ( d x^\mu) &=& \varepsilon^{\mu\nu} (d x_{\nu}), \qquad
\tilde *\;\; ( d \theta) = (d \bar \theta), \qquad
\tilde *\;\; (d \bar \theta) = ( d \theta), \nonumber\\
\tilde *\; (d x^\mu \wedge d x^\nu) &=& \varepsilon^{\mu\nu}, \;\quad
\tilde * \;(d x^\mu \wedge d \theta) = \varepsilon^{\mu\theta}, \qquad
\tilde * \;(d x^\mu \wedge d \bar \theta) = \varepsilon^{\mu\bar \theta}, 
\nonumber\\
\tilde *\; (d \theta \wedge d \theta) &=& s^{\theta\theta}, \qquad
\tilde *\; (d \theta \wedge d \bar \theta) = s^{\theta \bar \theta}, \qquad
\tilde *\; (d \bar \theta \wedge d \bar \theta) = s^{\bar \theta \bar \theta}, 
\end{array} \eqno(4.1)
$$
where $ \varepsilon^{\mu \theta} = - \varepsilon^{\theta \mu},
\varepsilon^{\mu \bar \theta} = - \varepsilon^{\bar \theta \mu}$
and $ s^{\theta \bar \theta} = s^{\bar \theta\theta}$ etc. It is obvious that
the operation $(\tilde \delta \tilde A)$ would result in a superscalar
(zero-form) superfield (as $\tilde \delta$ reduces the degree of a super form
by one on which it operates). The explicit expression for this superfield
is
$$
\begin{array}{lcl}
\tilde \delta \tilde A &=& (\partial \cdot B) + s^{\theta\theta} 
(\partial_{\theta} \Phi) + s^{\bar \theta \bar\theta} (\partial_{\bar\theta}
\bar \Phi) + s^{\theta \bar \theta} (\partial_{\theta} \bar \Phi +
\partial_{\bar \theta} \Phi)\nonumber\\
&-& \varepsilon^{\mu \theta} (\partial_{\mu} \Phi + \varepsilon_{\mu\nu} 
\partial_{\theta} B^\nu) - \varepsilon^{\mu \bar \theta} (\partial_{\mu}
\bar \Phi + \varepsilon_{\mu\nu} \partial_{\bar \theta} B^\nu).
\end{array} \eqno(4.2)
$$
The analogue of the horizontality condition with the super co-exterior
derivative $\tilde \delta$ is to equate Eqn. (4.2) to the gauge-fixing term
$\delta A = (\partial \cdot A)$ (i.e., $ \tilde \delta \tilde A = \delta A)$. 
This restriction leads to the following conditions on the superfields
$$
\begin{array}{lcl}
\partial_{\theta} \bar \Phi + \partial_{\bar \theta} \Phi &=& 0, \quad\;\;
\partial_{\theta} \Phi = 0, \quad\;\;  
\partial_{\bar \theta} \bar \Phi = 0, \nonumber\\  
\partial_{\mu} \Phi + \varepsilon_{\mu\nu} \partial_{\theta} B^\nu &=& 0,
\qquad \;\;  \partial_{\mu} \bar \Phi 
+ \varepsilon_{\mu\nu} \partial_{\bar \theta} B^\nu = 0,
\end{array} \eqno(4.3)
$$
and an additional restriction on the local field components of the expansion
(3.4) for the bosonic superfield $B_{\mu} (x, \theta, \bar \theta)$. The latter
conditions are
$$
\begin{array}{lcl}
\partial \cdot \bar R = 0, \qquad \partial \cdot R = 0, \qquad
\partial \cdot S = 0.
\end{array} \eqno(4.4)
$$
The solutions for the restriction (4.3) are listed below
$$
\begin{array}{lcl}
R_{\mu}\; (x) &=& - \;\varepsilon_{\mu\nu}\; \partial^\nu \;\bar C (x), \quad
\bar R_{\mu}\; (x) = -\; \varepsilon_{\mu\nu}\; \partial^\nu \; C (x), \quad
\bar s\; (x) = 0, \nonumber\\
S_{\mu}\; (x) &=& +\; \varepsilon_{\mu\nu}\; \partial^\nu\;  E (x), 
\;\;\qquad\;\; s \;(x) = 0, \;\;\qquad \;\;(\partial \cdot A)\; (x) = 0,
\end{array} \eqno(4.5)
$$
which automatically satisfy the restrictions (4.4). It will be noticed that
the choice of the signs in the expansion(3.4) are such that the restriction
$ \partial_{\theta} \bar \Phi + \partial_{\bar \theta} \Phi = 0$ is
satisfied trivially.

In terms of solutions (4.5), the expansion (3.4) can be re-expressed as
$$
\begin{array}{lcl}
B_{\mu}\; (x, \theta, \bar \theta) &=& A_{\mu} (x) 
- \;\theta\; \varepsilon_{\mu\nu}\;\partial^{\nu}  C (x) 
- \;\bar \theta\;\varepsilon_{\mu\nu}\; \partial^{\nu} \bar C (x) 
+ \;i \;\theta \;\bar \theta \;
\varepsilon_{\mu\nu}\;\partial^{\nu}\; E (x), \nonumber\\
\Phi\; (x, \theta, \bar \theta) &=& C (x) 
- \;i \bar \theta\; E (x), \nonumber\\
\bar \Phi\; (x, \theta, \bar \theta) &=& \bar C (x) 
+\; i \; \theta\; E (x).
\end{array} \eqno(4.6)
$$
It is worth pointing out that the above expansion can be directly obtained from
the definition of $*$ operation in Section 2 (cf. Eqn. (2.6)). Now exploiting
dual- and anti-dual BRST symmetries (discussed in Sec. 2), we can rewrite
Eqn. (4.6) as
$$
\begin{array}{lcl}
B_{\mu}\; (x, \theta, \bar \theta) &=& A_{\mu} (x) 
+ \;\theta\; (\bar s_{d} A_{\mu} (x)) + \;\bar \theta\; (s_{d} A_{\mu} (x)) 
+ \;\theta \;\bar \theta (s_{d} \bar s_{d} A_{\mu} (x)), \nonumber\\
\Phi\; (x, \theta, \bar \theta) &=& C (x) + \; \bar \theta\; (s_{d} C (x)),
\nonumber\\ \bar \Phi\; (x, \theta, \bar \theta) &=& \bar C (x) 
+ \theta\; (\bar s_{d} \bar C (x)),
\end{array} \eqno(4.7)
$$
which is the analogue of Eqn. (3.10) of the previous section. We summarize this
section with the following comments: (i) (anti-) co-BRST symmetry
transformations are generated along the $\theta$- and $\bar \theta$ directions
of the supermanifold. (ii) The translation generators along the Grassmannian
directions of the supermanifold are the conserved and nilpotent
(anti-)co-BRST charges. (iii) For the odd (fermionic) superfields, the
translations are either along $\theta$ or $\bar \theta$ directions (unlike
the bosonic superfield where translations are along both $\theta$ as well
as $\bar \theta$ directions). (iv) Comparison between (3.10) and (4.7) shows
that the (anti-)BRST transformations are along $(\theta)\bar \theta$
directions for the odd fields $(C)\bar C$. {\it On the contrary, the (anti-) 
co-BRST transformations are the other way around}. (v) A single restriction 
$ \tilde \delta \tilde A = \delta A$ 
produces co-BRST- and anti-co-BRST symmetry transformations for the Lagrangian
density (2.1). Thus, the mapping is: $ \tilde \delta \Leftrightarrow (Q_{d},
\bar Q_{d})$ but the ordinary co-exterior derivative $\delta$ is identified with
$Q_{d}$ {\it alone} because it decreases the ghost number of a state by one
[11,12,17] as $\delta$ reduces the degree of a given form by one on which it 
operates [7-10].\\

\noindent
{\bf 5 Discrete symmetries}\\

\noindent
We have discussed a few discrete symmetries at the fag end of Sec. 2. Now we
exploit these discrete symmetries vis-a-vis our superfield expansion (3.4). We
emphasize the fact that, for the BRST- and dual BRST symmetries, we have
shown that: $ s(x) = 0, \bar s(x) = 0$ in the expansion (3.4). Thus, we
shall now be concentrating on (3.4) {\it only for this case}.  First of all, 
it is straightforward to verify that under the following discrete 
transformations
$$
\begin{array}{lcl}
&&C\rightarrow \pm i \bar C,\quad \bar C \rightarrow \pm i C, \quad
E \rightarrow \pm i (\partial \cdot A),\quad (\partial \cdot A) \rightarrow 
\pm i E, \quad 
\partial_{\mu} \rightarrow \pm i
\varepsilon_{\mu\nu} \partial^\nu,\nonumber\\
&& \theta \rightarrow - \theta, \quad \bar \theta \rightarrow - \bar \theta,
\quad R_{\mu} \rightarrow - R_{\mu}, \quad\; \bar R_{\mu} \rightarrow - \bar
R_{\mu}, \quad S_{\mu} \rightarrow S_\mu,\quad\;
A_{\mu} \rightarrow A_{\mu}, 
\end{array} \eqno(5.1)
$$
the superfields in (3.4) undergo the following change
$$
\begin{array}{lcl}
\Phi (x, \theta, \bar \theta) \rightarrow \pm \; i\; \bar \Phi (x, \theta,
\bar \theta),\;\;
\bar \Phi (x, \theta, \bar \theta) \rightarrow \pm \; i\; \Phi (x, \theta,
\bar \theta),\;\;
B_{\mu} (x, \theta, \bar \theta) \rightarrow
B_{\mu} (x, \theta, \bar \theta). 
\end{array} \eqno(5.2)
$$
Furthermore, it can be trivially checked that the above 
transformations still satisfy:
$ \Phi^2 = 0, \bar \Phi^ 2 = 0, \Phi \bar \Phi + \bar \Phi \Phi = 0$ and 
$ [ B_{\mu}, B_{\nu} ] = 0$. Yet another interesting point is to see that
in the limit: $ \theta \rightarrow 0,\;
\bar \theta \rightarrow 0$, the above transformations reduce to
: $ C \rightarrow \pm i \bar C, \bar C \rightarrow \pm i C,
A_{\mu} \rightarrow A_{\mu}$. Thus, transformations (5.2) are the 
generalization of the discrete symmetry (2.6).  A close look at the 
expressions for $R_{\mu}, \bar R_{\mu}, S_{\mu}$ in equations
(3.9) and (4.5) allows us to write down the Hodge decomposed versions for
these 2D fermionic ($R_{\mu}, \bar R_{\mu}$)- and bosonic ($S_{\mu}$) vectors
(appearing in the expansion of the
bosonic superfield $B_{\mu} (x, \theta, \bar \theta)$) as
$$
\begin{array}{lcl}
R_{\mu} = \partial_{\mu} C + \varepsilon_{\mu\nu} \partial^\nu \bar C,\quad
\bar R_{\mu} = \partial_{\mu} \bar C + \varepsilon_{\mu\nu} \partial^\nu C,
\quad S_{\mu} = + \partial_{\mu}\;(\partial \cdot A)  
- \varepsilon_{\mu\nu} \partial^\nu E,
\end{array} \eqno(5.3)
$$
which are solutions to the transformations: $ R_{\mu} \rightarrow - R_{\mu},
\bar R_{\mu} \rightarrow - \bar R_{\mu}, S_{\mu} \rightarrow S_{\mu}$ under
the discrete transformations (2.6). However, it is interesting to note that 
the r.h.s. of
the expression for $S_{\mu}$ is the equation of motion for the 2D 
photon: $\partial_{\mu} F^{\mu\nu} + \partial^\nu (\partial \cdot A) = 0$
(with $F^{10} = E$). Thus, $S_{\mu}$ turns out to be zero on the on-shell.
{\it It can be checked that $S_\mu = - \varepsilon_{\mu\nu} \Box A^\nu$ 
also transforms as $S_{\mu}\rightarrow S_\mu$ under (2.6) because
$\Box \rightarrow \Box$ under $\partial_\mu \rightarrow \pm i 
\varepsilon_{\mu\nu} \partial^\nu$}. 
Now let us concentrate on the discrete symmetries: $ C \leftrightarrow \bar
C, E \leftrightarrow - E, (\partial \cdot A) \leftrightarrow - (\partial \cdot
A)$ that connect BRST- to anti-BRST- as well as co-BRST- to anti-co-BRST 
symmetry transformations. The generalized version of these symmetries,
vis-a-vis our superfield expansion (3.4), is:
$$
\begin{array}{lcl}
&& C \leftrightarrow \bar C, \qquad
(\partial \cdot A) \leftrightarrow - (\partial \cdot A), \qquad
E \leftrightarrow - E, \nonumber\\
&& \theta \leftrightarrow \bar \theta, \quad
R_{\mu} \leftrightarrow - \bar R_\mu, \quad
\bar R_{\mu} \leftrightarrow - R_{\mu}, \quad
S_{\mu} \leftrightarrow S_{\mu}, 
\end{array} \eqno(5.4)
$$
under which the superfields transform as
$$
\begin{array}{lcl}
\Phi \;\leftrightarrow\; \bar \Phi,  \quad
(\partial \cdot B)\; \leftrightarrow\; -\; (\partial \cdot B), \quad
- \varepsilon^{\mu\nu} \partial_{\mu} B_{\nu}\;
\leftrightarrow \;
 \varepsilon^{\mu\nu} \partial_{\mu} B_{\nu}.
\end{array} \eqno(5.5)
$$
It will be noticed that in the limit $\theta \rightarrow 0, \bar \theta
\rightarrow 0$, we get back our original discrete symmetries: $ C
\leftrightarrow \bar C, (\partial \cdot A) \leftrightarrow 
- (\partial \cdot A), E \leftrightarrow - E$. It is interesting to point out
that the solutions (5.3) are no longer the appropriate solutions
for the present case. In fact, taking the help of (3.9) and (4.5), now the
solutions for the 2D fermionic vectors are
$$
\begin{array}{lcl}
R_{\mu} = \partial_{\mu} C - \varepsilon_{\mu\nu} \partial^\nu \bar C, \qquad
\bar R_{\mu} = - \partial_{\mu} \bar C + \varepsilon_{\mu\nu} \partial^\nu C,
\end{array} \eqno(5.6)
$$
which are nothing but the {\it orthogonal} Hodge decomposed version of 
the corresponding solution in (5.3). Now, for the present case where
$(\partial \cdot A) \leftrightarrow - (\partial \cdot A), \;
E \leftrightarrow - E$, it is clear that any arbitrary linear combination:
$ S_{\mu} = P\; \partial_{\mu} (\partial \cdot A) + Q\; \varepsilon_{\mu\nu}
\partial^\nu E$ (where $P$ and $Q$ are some c-number constants)
would lead to $S_{\mu} = 0$ for the requirement $ S_{\mu} \rightarrow 
S_{\mu}$ (cf. (5.4)) to be satisfied.  {\it The origin for the
existence of the (anti-)BRST- and (anti-)co-BRST symmetries in the theory
is encoded in the orthogonal relations (5.3) and (5.6) for the Hodge
decomposed versions of $R_{\mu}$ and $\bar R_{\mu}$.} In fact, these relations
show that $\partial_{\mu} C (\partial_\mu \bar C)$ and $\varepsilon_{\mu\nu}
\partial^\nu \bar C (\varepsilon_{\mu\nu} \partial^\nu C)$ are the separate 
and independent symmetry transformations for the Lagrangian density (2.1). In
the language of the BRST cohomology and HDT, this is the logical explanation for
the existence of (anti-)BRST- and (anti-)co-BRST symmetries for the Lagrangian
density (2.1) of a free 2D Abelian gauge theory.\\

\noindent
{\bf 6 Super Laplacian operator and bosonic symmetry}\\

\noindent
For the sake of brevity, we shall consider the
expansion (3.4) for the case $ s(x) = \bar s(x) = 0$. The analogue of the
horizontality condition w.r.t. super Laplacian operator $\tilde \Delta$ is
$$
\begin{array}{lcl}
\tilde \Delta \; \tilde A = \Delta \;A, \quad
\tilde \Delta  = \tilde d \tilde \delta + \tilde \delta \tilde d, \quad
\Delta  =  d  \delta +  \delta  d. 
\end{array} \eqno(6.1)
$$
It is obvious that $\Delta A = d x^\mu\; [\; \partial_\mu (\partial \cdot A)
- \varepsilon_{\mu\nu} \partial^\nu E \;] = d x^\mu\; \Box A_\mu$. Now we can
check that the l.h.s. of (6.1) (with $\tilde \delta = - \tilde *\;\tilde d
\; \tilde *$) can be rewritten as
$$
\begin{array}{lcl}
\tilde d\; (\tilde \delta\; \tilde A) &=& 
d x^\rho\; \partial_{\rho}\; (\tilde \delta \tilde A)\;
+\; d \theta\; \partial_{\theta}\; (\tilde \delta \tilde A)\;
+ \;d \bar \theta\; \partial_{\bar \theta}\; (\tilde \delta \tilde A), 
\nonumber\\
\tilde \delta\; (\tilde d \;\tilde A) &=& 
d x^\rho\; \varepsilon_{\rho\lambda}\; \partial^{\lambda} 
\;[\tilde * (\tilde d \tilde A)] - \;d \theta\; \partial_{\bar \theta} \;
[\tilde * (\tilde d \tilde A)] - \;d \bar \theta\; \partial_{\theta} \; 
[\tilde * (\tilde d \tilde A)],
\end{array} \eqno(6.2)
$$
where the explicit expression for the term in the square bracket is
$$
\begin{array}{lcl}
\tilde * (\tilde d \tilde A) &=& \varepsilon^{\mu\nu} \partial_{\mu} B_{\nu}
+ \varepsilon^{\mu \theta} (\partial_{\mu} \bar \Phi - \partial_{\theta}
B_\mu) + \varepsilon^{\mu\bar\theta}
(\partial_{\mu}  \Phi - \partial_{\bar \theta} B_{\mu}) - s^{\theta\theta}
(\partial_{\theta} \bar \Phi) \nonumber\\
&-& s^{\bar\theta\bar\theta} (\partial_{\bar \theta} \Phi)
- s^{\theta\bar\theta} (\partial_{\bar \theta} \bar \Phi + \partial_{\theta}
\Phi).
\end{array} \eqno(6.3)
$$
Equation (6.1) can be expressed in a more transparent way as follows
$$
\begin{array}{lcl}
d x^\rho \bigl [\; \partial_{\rho} \;(\tilde \delta \tilde A) + \;
\varepsilon_{\rho \lambda}\; \partial^\lambda\; 
\{ \tilde * (\tilde d \tilde A)\}
\;\bigr ] = d x^\rho\; \Box\; A_\rho, \end{array} \eqno(6.4)
$$
$$
\begin{array}{lcl}
d \theta\; \bigl [\; \partial_{\theta}\; (\tilde \delta \tilde A) -
\;\partial_{\bar \theta} \;\{ \tilde * (\tilde d \tilde A)\}
\;\bigr ] = 0, \end{array} \eqno(6.5)
$$
$$
\begin{array}{lcl}
d \bar \theta\; \bigl [\; \partial_{\bar \theta}\; (\tilde \delta \tilde A) -
\;\partial_\theta \;\{ \tilde * (\tilde d \tilde A)\}
\;\bigr ] = 0. \end{array} \eqno(6.6)
$$
The last requirement in the above equation leads to the following restrictions
$$
\begin{array}{lcl}
\partial \cdot S = 0, \;\; \varepsilon^{\mu\nu} \partial_{\mu} S_{\nu} = 0,
\quad \partial \cdot R = \varepsilon^{\mu\nu} \partial_\mu \bar R_\nu, \quad
S_\mu = - \partial_\mu (\partial \cdot A), \; S_\mu =  \varepsilon_{\mu\nu}
\partial^\nu E.
\end{array} \eqno(6.7)
$$
It is clear that $ R_\mu = \varepsilon_{\mu\nu} \bar R^\nu$ and
the two expressions for  $S_\mu$ lead to
$$
\begin{array}{lcl}
S_\mu = - \frac{1}{2}\; \bigl [ \partial_\mu (\partial \cdot A) 
- \varepsilon_{\mu\nu} \partial^\nu E \bigr ], \quad
\partial_{\mu} (\partial \cdot A) + \varepsilon_{\mu\nu} \partial^\nu E = 0,
\end{array} \eqno(6.8)
$$
where the r.h.s. of $S_\mu$ is nothing but the equation of motion for the
2D free photon. The latter equation is not invariant
under the ``duality'' transformations (2.6) 
and $R_\mu = \varepsilon_{\mu\nu} \bar R^\nu$ is satisfied for the 
solutions (5.3) as well as (5.6). The condition (6.5) leads to
$$
\begin{array}{lcl}
\partial \cdot S = 0, \;\; \varepsilon^{\mu\nu} \partial_{\mu} S_{\nu} = 0,
\quad \partial \cdot \bar R = \varepsilon^{\mu\nu} \partial_\mu R_\nu, \quad
S_\mu = - \partial_\mu (\partial \cdot A), \; S_\mu =  \varepsilon_{\mu\nu}
\partial^\nu E.
\end{array} \eqno(6.9)
$$
It is evident that now $\bar R_\mu = \varepsilon_{\mu\nu}  R^\nu$ and
the two expressions for $S_\mu$ lead to the same conclusions as in (6.8).
In fact, equation (6.8) implies that all the conditions on $S_\mu$ (i.e.
$\partial \cdot S = 0, \;\varepsilon^{\mu\nu} \partial_{\mu} S_{\nu} = 0,
\;\Box S_\mu = 0$) are satisfied because $\Box (\partial \cdot A) = 0$
and $\Box E = 0$. 
The consistency with the equation of motion, however, implies that $S_\mu = 0$ 
on the on-shell. Furthermore, the requirement of duality invariance of the 
latter equation in (6.8) 
forces us to choose: $\partial_\mu (\partial \cdot A) = 0,\;
\varepsilon_{\mu\nu} \partial^\nu E = 0$. As an operator equation, the more
stringent restrictions:   
 $(\partial \cdot A) = 0$ and $E = 0$ are expected because if we
choose the harmonic states to be the physical state of the theory then 
$Q_{b} |phys> = 0 $ (with
$Q_{b} = \int dx [\partial_{0} (\partial \cdot A) C - (\partial \cdot A) 
\dot C]$) and $Q_{d} |phys> = 0$ (with $Q_{d} = \int dx [ E \dot {\bar C}
- \dot E \bar C]$) imply that $(\partial \cdot A) |phys> = 0$ and
$ E |phys> = 0$ [11,12,17]. Now, Eqn.(6.4) yields the relations:
$ d x^\rho\; \Box B_{\rho} = d x^\rho\; \Box A_\rho \Leftrightarrow
\Box R_{\rho} = \Box \bar R_\rho = \Box S_\rho = 0.  $
Setting the coefficients of $(d x^\rho s^{\theta\theta}), (d x^\rho 
s^{\bar\theta \bar \theta})$ equal to zero leads to: $\partial_\rho (\partial
\cdot A) - \varepsilon_{\rho\lambda} \partial^\lambda E = 0$ which,
once again,  establishes the fact that $S_\mu = 0$ in (6.8). The operator
equations: $(\partial \cdot A) = 0, E = 0$ also imply the same. 
Note that the coefficient
of $(d x^\rho s^{\theta \bar \theta})$ leads to no new restrictions as choice 
of signs in the expansion (3.4) satisfies it trivially. Lastly, setting 
the coefficients of $(d x^\rho \varepsilon^{\mu\theta})$ and 
$(d x^\rho \varepsilon^{\mu\bar\theta})$ equal to zero leads to
$$
\begin{array}{lcl}
\partial_\mu (\varepsilon_{\rho\lambda} \partial^\lambda C - \partial_\rho
\bar C) &=& \varepsilon_{\rho\lambda} \partial^\lambda R_\mu + \partial_\rho
(\varepsilon_{\mu\nu} R^\nu), \nonumber\\
\partial_\mu (\varepsilon_{\rho\lambda} \partial^\lambda \bar C - \partial_\rho
C) &=& \varepsilon_{\rho\lambda} \partial^\lambda \bar R_\mu + \partial_\rho
(\varepsilon_{\mu\nu} \bar R^\nu), \nonumber\\
\partial_\mu [\varepsilon_{\rho\lambda} \partial^\lambda 
(\partial \cdot A) - \partial_\rho
E ] &=& - [\varepsilon_{\rho\lambda} \partial^\lambda S_\mu + \partial_\rho
(\varepsilon_{\mu\nu} S^\nu ]. 
\end{array} \eqno(6.10)
$$
The last equation is satisfied due to $(\partial \cdot
A) = 0, E = 0, \partial \cdot S = 0, \varepsilon^{\mu\nu} \partial_\mu S_\nu
= 0$.  It is clear that for $\partial \cdot A = 0, E = 0$, we obtain 
$S_\mu = 0$ in (6.8). However, there
is another choice $S_\mu = - \varepsilon_{\mu\nu} \Box A^\nu$ that remains 
invariant under both the discrete symmetries (5.1) and (5.4) 
but vanishes on the on-shell ($\Box A_\mu = 0$). The other two coupled 
equations for the fermionic vectors (with $R_\mu = \varepsilon_{\mu\nu} 
\bar R^\nu, \bar R_\mu = \varepsilon_{\mu\nu} R^\nu$) are satisfied for the 
choice of Hodge decomposed versions (5.6) with the restrictions $R_\mu 
= \bar R_\mu = 0$. More precisely, these equations lead to: $R_\mu 
= \partial_\mu C - \varepsilon_{\mu\nu} \partial^\nu \bar C = 0$
and $\partial_\mu R_\nu + \varepsilon_{\mu\lambda} \partial^\lambda \bar R_\nu
= 0.$ Thus, ultimately, we have obtained:
$ R_{\mu} = 0,  \bar R_\mu = 0,  S_\mu = - \varepsilon_{\mu\nu}
\Box A^\nu,  \partial \cdot A = 0,  E = 0.$ With these values together with 
$ s(x) = \bar s(x) = 0$ and the observation that ($s_{w} A_\mu = 
- \varepsilon_{\mu\nu} \Box A^\nu$), we have expansion (3.4) as
$$
\begin{array}{lcl}
B_\mu (x,\theta,\bar\theta) = A_\mu (x) + i \theta \bar \theta 
(s_{w} A_\mu (x)), \quad \Phi (x,\theta,\bar \theta) = C (x), \quad
\bar \Phi (x,\theta,\bar \theta) = \bar C (x),
\end{array} \eqno(6.11)
$$
which shows that there are no transformations for the (anti-)ghost fields 
but the gauge field $A_\mu$ {\it alone} transforms to its own equation of 
motion (cf. Sec. 2) along the ($\theta\bar\theta$) direction.\\

\noindent
{\bf 7 Conclusions}\\

\noindent
We have demonstrated the existence of some new local symmetries by exploiting 
the mathematical power of the super de Rham cohomology operators of 
differential geometry defined
on a $(2 + 2)$-dimensional compact supermanifold. As conserved and nilpotent 
(anti-)BRST charges $(\bar Q_{b})Q_{b}$ are connected with the super
exterior derivative $\tilde d$ [3,4], in a similar fashion (anti-)co-BRST 
charges $(\bar Q_{d})Q_{d}$ are connected with the super 
co-exterior derivative $\tilde \delta$. These nilpotent charges turn out to
be the translation generators along the Grassmannian directions of the
supermanifold. A bosonic charge $Q_{w}$ is shown to be related with the super
Laplacian operator $\tilde \Delta$. This charge turns out to be the
translation generator along the bosonic direction (which is equivalent to a
couple of intertwined Grassmannian directions) of the supermanifold. The mapping
between super operators $(\tilde d, \tilde \delta, \tilde \Delta)$ and the local
conserved charges is: $\tilde d \Leftrightarrow (Q_{b}, \bar Q_{b}), 
\tilde \delta
\Leftrightarrow (Q_{d}, \bar Q_{d}), \tilde \Delta \Leftrightarrow Q_{w}$. 
The analogy between the ghost number of a
state in the quantum Hilbert space and the degree of a differential form
allows one to relate the ordinary de Rham cohomology operators $(d, \delta,
\Delta)$ with the conserved charges as: $ d \Leftrightarrow (Q_{b}, \bar Q_{d}),
\delta \Leftrightarrow (Q_{d}, \bar Q_{b}), \Delta \Leftrightarrow Q_{w}
= \{ Q_{b}, Q_{d} \} = \{ \bar Q_{b}, \bar Q_{d} \}$. In the setting of the
superfield formulation, the above mappings find their geometrical 
interpretation. The interplay between the discrete- and continuous symmetries
of the theory allows one to write down the Hodge decomposed versions for the
2D fermionic vectors which provide an unambiguous explanation for the existence
of (anti-)BRST- and (anti-)co-BRST symmetries in the theory. It would be nice
to extend these ideas to the interacting case [14,15].

\end{document}